# Transformation Pathways of Silica under High Pressure

Liping Huang, Murat Durandurdu, and John Kieffer[*]

*Department of Materials Science and Engineering, University of Michigan, Ann Arbor, Michigan 48109-2136, USA*

Concurrent molecular dynamics simulations and *ab initio* calculations show that densification of silica under pressure follows a ubiquitous two-stage mechanism. First, anions form a close-packed sub-lattice, governed by the strong repulsion between them. Next, cations redistribute onto the interstices. In cristobalite silica, the first stage is manifest by the formation of a metastable phase, which was observed experimentally a decade ago, but never indexed due to ambiguous diffraction patterns. Our simulations conclusively reveal its structure and its role in the densification of silica.

Silica is one of the most abundant minerals in the Earth's crust and an important engineering material. Knowledge of the densification mechanisms in this material is important for elucidating questions ranging from the constitution of the transition zone between the Earth's upper and lower mantle [1-3], to amorphization upon impact [4, 5], or developing high-toughness ceramics [6]. Because direct examination of matter under such severe loading conditions is often impossible, researchers must resort to alternative methods for investigating these processes. Using atomic-scale simulations, Tsuneyuki *et al*. [7] predicted that *α*-cristobalite transforms into a metastable phase at 16.5 GPa. In subsequent x-ray diffraction experiments, Tsuchida *et al*. [8] indeed observed that compressed *α*-cristobalite transforms into several unknown phases labeled X-I, X-II and X-III. However, none of their x-ray diffraction patterns matched that of the predicted phase, or any known silica polymorph. Attempts to index these structures have so far failed because of the weak and broad diffraction peaks. Based on molecular dynamics (MD) simulations, using a new reactive force field, and *ab initio* calculations, we are now able to not only index the structure of the X-I phase, but furthermore reveal its significance for the densification of silica as the first stage during the transformation from cristobalite to stishovite. The transformation pathway we discovered is remarkable in that the stability ranges of stishovite and cristobalite do not adjoin, and the possibility of a direct transformation between the two polymorphs is not indicated.

---

[*]Electronic mail: kieffer@umich.edu

We performed the first-principles calculations within the density-functional theory formalism, using the generalized-gradient approximation (GGA) [9] for the exchange-correlation energy. The calculations were carried out with the SIESTA program [10] using a linear combination of atomic orbitals as the basis set, and norm-conserving Troullier-Martins pseudopotentials [11]. An optimized split-valence double-$\zeta$ plus polarized basis set was employed and the plane wave cutoff was at 120 Ry. Γ-point sampling for the super cell's Brillouin zone integration was used, which is reasonable for a cell with 96 atoms. For each value of the applied stresses, the lattice vectors were optimized together with the atomic coordinates by using the conjugate-gradient technique.

MD simulations were carried out on the systems of 64, 125 and 216 unit cells of $\alpha$-cristobalite (12 atoms per cell) and 64 and 216 unit cells of $\alpha$-quartz (9 atoms per cell) with periodic boundary conditions to ascertain the absence of system size effect. A coordination-dependent charge transfer three-body potential [12] was employed in this study, using one single force field parameterization for all the simulated phases. The hydrostatic pressure is applied to each system in steps of 1 GPa followed by a 20 ps equilibration period.

Our simulations show that upon densification, both cristobalite and quartz follow an analogous two-step process. A similar mechanism has been proposed earlier to account for the coordination changes in silicate melts, glasses and possibly pressure-induced amorphization [13]. In presenting the characteristics of this process we will concentrate on cristobalite, as it involves the added interest of disclosing the nature of a so far unidentified high-pressure phase. We also discuss our results for quartz. For one, it validates the simulation approach through comparison with the work done by others [14], and second, it corroborates the notion of universality for this two-step process as the densification mechanism in oxides.

$\alpha$-cristobalite has a tetragonal $P4_12_12$ symmetry under ambient conditions. The unit cell is defined by two lattice constants ($a$=4.9733 Å, $c$=6.9262 Å) and four internal parameters for the atomic positions ($u$=0.3 for Si, and $x$=0.245, $y$=0.10, $z$=0.175 for O). In our MD simulations, $\alpha$-cristobalite is stable up to 15 GPa (Fig. 1(a)), and then undergoes a structural transition. This pressure is very close to that at which the X-I phase emerges in experiments [15-20]. Both *ab initio* and MD simulations predict the same transformation pathway. Discrepancies in the density vs. pressure behavior can be attributed in part to the fact that *ab initio* calculations are carried out at zero Kelvin and MD simulations at finite temperatures, and in part to the differences in the pressure control algorithms and system sizes used in *ab initio* vs. MD simulations. After going through an intermediate state with orthorhombic $C222_1$ symmetry between 15 and 17 GPa, the structure regains the $P4_12_12$ symmetry (determined by using KPLOT [21]). The unit cell at 20 GPa and 300 K has different parameters ($a$=4.3552 Å, $c$=5.9599 Å, $u$=0.3803, $x$=0.2310, $y$=0.2105, $z$=0.2044). This four-coordinated high-pressure phase (hp-cristobalite, for short) is unquenchable, i.e., upon decompression it reverts to α-cristobalite following the same path with negligible hysteresis. Upon further compression, hp-cristobalite transforms into six-coordinated stishovite at ~22 GPa, which is associated with the density jump seen in Fig. 1(a). In the absence of thermal motion this transition is shifted to higher pressure, and in our *ab initio* calculations it occurs at ~36 GPa. A simulated stishovite configuration at 25 GPa is shown in Fig. 1(b); it has $P42/mnm$ symmetry ($a$=4.1596 Å, $c$=2.8365 Å, $u$=0.0, $x$=0.304, $y$=0.304, $z$=0.0) at 300 K. At ~40 GPa, this structure displacively transforms into the post-stishovite $CaCl_2$ phase by rotating the $SiO_6$ octahedra relative to each other about the $c$ axis (Fig. 1(b)). The unit cell changes from tetragonal to orthorhombic ($a \neq b$, $x \neq y$), as postulated from theoretical considerations [2, 22-24]. In experiments, this transformation occurs at pressures above 45−50 GPa at ambient temperature [1-3]. Upon releasing pressure, this phase reproducibly reverts back to stishovite, and as in experiments, our simulated stishovite can be quenched down to ambient pressure without coordination reversion.

Finally, while the experimental x-ray diffraction pattern of the X-I phase has weak and broad features, peaks become much sharper when this phase is heated, and eventually the diffraction pattern converts into that of stishovite [8, 16]. Again, our simulated hp-cristobalite also exhibits this behavior: when thermally activated it transforms into stishovite without need for further compression. Overall, our simulations reproduce the experimentally observed transformation pathways from cristobalite to post-stishovite remarkably well. This provides credence to the accuracy of our interaction model and the ensuing analysis of the structures we generated, particularly the hp-cristobalite phase and its relation to the X-I phase.

In the x-ray diffraction pattern of the X-I phase, which has been determined by a number of investigators [8, 16, 18, 19], it is impossible to explain the diffraction line at a $d$-spacing of ~3.5 Å by any known polymorphs of silica, even when assuming asymmetrically distorted unit cells. An excellent agreement is found between the diffraction pattern of the X-I phase and that of the simulated hp-cristobalite structure at 20 GPa (Fig. 2(a)). The small deviation in the peak positions around 2.7 Å may be due to the different pressure conditions, i.e., in experiments hydrostaticity is not perfect whereas in simulations it is. The simulated x-ray diffraction pattern for stishovite is also in excellent agreement with experimental data [25].

Figure 2(b) shows the pressure dependence of the simulated infrared spectra. With increasing pressure, the strong Si-O-Si asymmetric stretching mode at ~1100 cm$^{-1}$, which is characteristic of four-coordinated silicon, clearly splits due to the distortion of the SiO$_4$ tetrahedra. At 20 GPa, our simulated spectrum is in good agreement with that of the X-I phase, both in terms of peak positions and their relative intensities [17]. For stishovite, which consists only of six-coordinated Si, the Si-O-Si stretching mode gives rise to a peak doublet near 900 cm$^{-1}$ and no feature is observed near 1100 cm$^{-1}$.

From the above observations, we conclude that the hp-cristobalite phase in our simulations indeed corresponds to the experimentally observed X-I phase (or C-III, according to Hemley [26]). From the difference in the frequencies of Si-O-Si stretching modes in tetrahedral and octahedral bonding environment, we can establish that the X-I phase has only four-coordinated Si. This is consistent with the interpretation of experimental Raman spectra [18, 20], but in contrast to the earlier simulations of compressed cristobalite, which predicted a phase with *Cmcm* symmetry, containing an equal number of four- and six-coordinated Si [7]. The absence of the X-I to stishovite transformation at room temperature in experiments may be due to the lack of kinetic energy and/or non-hydrostatic conditions.

Next we assert the metastable nature of the X-I phase and explain why it is favored as a transient structure that readily transforms into other crystalline forms [8, 16] or amorphizes [4] under pressure. Figure 3 shows *ab initio* calculated total energies as a function of volume for *α*-cristobalite, the X-I phase, and stishovite. These energy-volume relationships uniquely delineate the stability ranges of the respective phases. The tangent common to pairs of curves defines the volumes at which phase would transform into one another. Due to topological incompatibilities, cristobalite does not simply transform into another stable polymorph, e.g., quartz or coesite, whose energy-volume curve would be intermediate to cristobalite and stishovite [27]. Instead, the configuration follows state points along the solid curve and eventually reaches the stability range of the X-I phase. The energy-volume curve of the latter is only slightly lower than that of *α*-cristobalite, and its minimum occurs at a volume at which the energy of α-cristobalite is in fact lower. The common tangent between the energy curves of α-cristobalite and X-I indicates a small density change upon transformation between these two phases, which explains why this transition occurs gradually and requires very little activation. On similar grounds, we expect the transition from the X-I phase to stishovite to be sharp and accompanied by a significant volume change, which is consistent with the density vs. pressure behavior shown in Fig. 1(a).

The role of the X-I phase in the overall densification process becomes evident when examining the transformation mechanisms. To this end we track the fractional atomic coordinates relative to α-cristobalite unit cell as a function of pressure (Fig. 4). Accordingly, Si atoms move along the diagonal to the center of the basal plane, while $SiO_4$ tetrahedra rotate about the [110] direction. The alignment of O atoms within {110} planes of the X-I phase, as evidenced by the converging $x$ and $y$ fractional coordinates, results in a distorted *hcp* anion lattice similar to that in the rutile structure (Fig. 1(b)). However, the emerging dominance of the anion lattice causes the $SiO_4$ tetrahedra to distort. The O-Si-O angles parallel to the basal plane open up to 130°, creating low-energy passageways for the Si atoms to move from fourfold to six-fold bonding. Thus, under further compression or thermal activation, Si atoms shift to the edge of $SiO_4$ tetrahedra ($u$ jumping to 0.5 in Fig. 4), while the positions of O atoms slightly adjust ($x$=0.304, $y$=0.304, $z$=0.25) to form 'perfect' $SiO_6$ octahedra (inset of Fig. 4). By the time the Si atoms achieve octahedral coordination, one unit cell of X-I has split into two unit cells of stishovite along the *c* axis, with $P4_2/mnm$ symmetry ($u$=0.0, $x$=0.304, $y$=0.304, $z$=0.0 after a change of origin). The transformation pathways from α-cristobalite to stishovite we observed are similar to the mechanisms proposed by O'Keefe *et al*. based on theoretical considerations [28], except that cations and anions respond to pressure in a sequential fashion rather than moving simultaneously. Anions form a close packing first, followed by sudden shift in cation positions that results in an abrupt coordination change.

Similarly, using MD simulations we observed the formation of a body centered cubic (*bcc*) anion sub-lattice in compressed α-quartz [29], thereby confirming predictions from first-principles calculations [30] and experimental observations [31]. The formation of close-packed anion sub-lattices under pressure, with cations distributed over part of the interstices, has been described as a eutaxic ordering [32], a process governed by the balance between attractive (unlike ions) and repulsive (like ions) forces, while simultaneously maximizing density. Our simulations show that such eutaxic arrangements are common to at least the two major polymorphs of silica under pressure. We therefore submit that this process may constitute a ubiquitous model for the densification under pressure of inorganic compounds with rigid polyhedral building blocks at ambient conditions. Depending on the synchronization of cations movement between interstices different transformations can be observed. In the extreme, when cations move completely independently from each other upon compression, amorphization takes place [4], while more cooperative displacements would be consistent with earlier theoretical explanations of the extensive polymorphism in silica [22]. In this scheme, the X-I phase and stishovite can be considered as an *hcp* anion lattice with cations located in tetrahedral or octahedral interstices, respectively. The significance of the X-I phase is that it establishes the anion packing that facilitates a low-activation pathway from cristobalite to stishovite.

L. H is grateful to R. Hundt for discussion on determining the symmetries of the simulated structures. This work was supported by National Institute of Standards and Technology and the National Science Foundation.

**Figure Captions**

FIG. 1. Density and structure evolution of cristobalite silica under pressure. (a) Density from MD simulations and *ab initio* calculations. (b) MD simulated structure viewed along the *c* axis at 0, 20, 25 and 40 GPa. The large spheres represent O atoms; the small ones Si atoms.

FIG. 2. Structural and dynamical property of the simulated structures at 0, 20 and 25 GPa. (a) X-ray diffraction patterns, comparing to that of X-I (shown as vertical ticks on a dotted line [8, 16, 18, 19]). (b) Infrared spectra.

FIG. 3. The calculated total energies as a function of volume for $\alpha$-cristobalite, X-I phase and stishovite. The energy of the $\alpha$-cristobalite at zero pressure is taken as the origin of energy. The lines are fitted to the data by second order polynomials.

FIG. 4. Fractional coordinates with respect to the $\alpha$-cristobalite unit cell as a function of pressure. Inset shows side view of the unit cell of $\alpha$-cristobalite, X-I and stishovite, respectively. The large spheres represent O atoms; the small ones Si atoms.

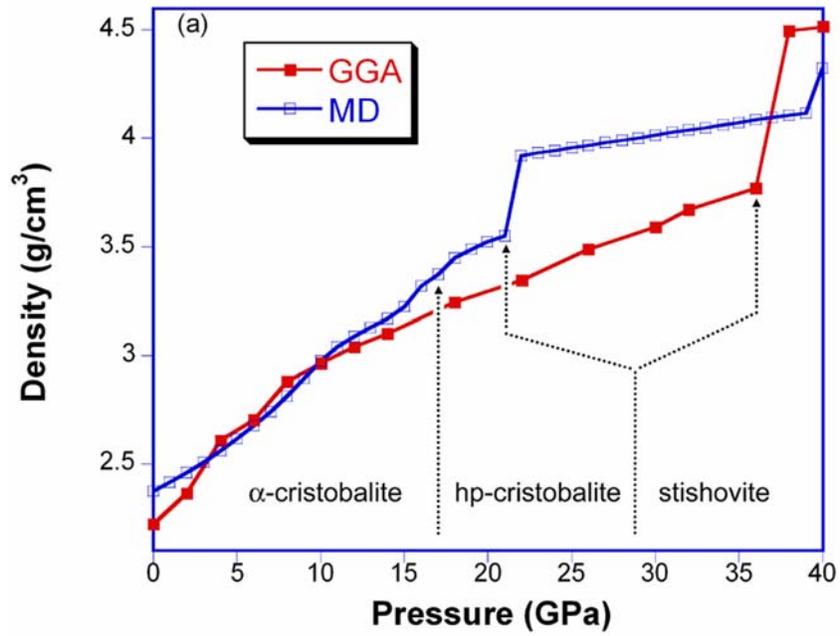

Fig. 1(a)-Huang-2005

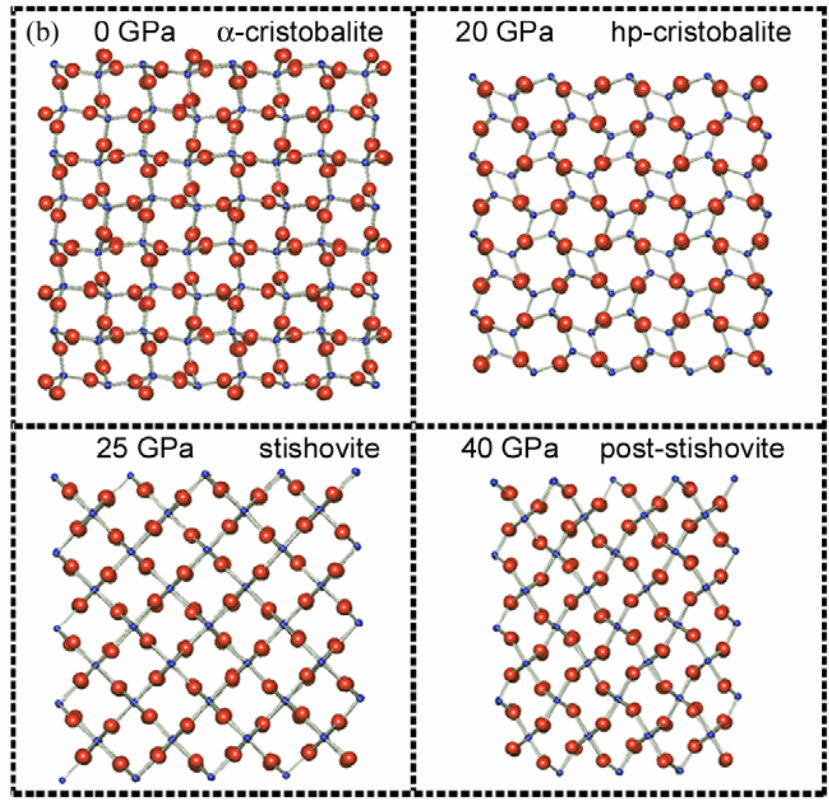

Fig. 1(b)-Huang-2005

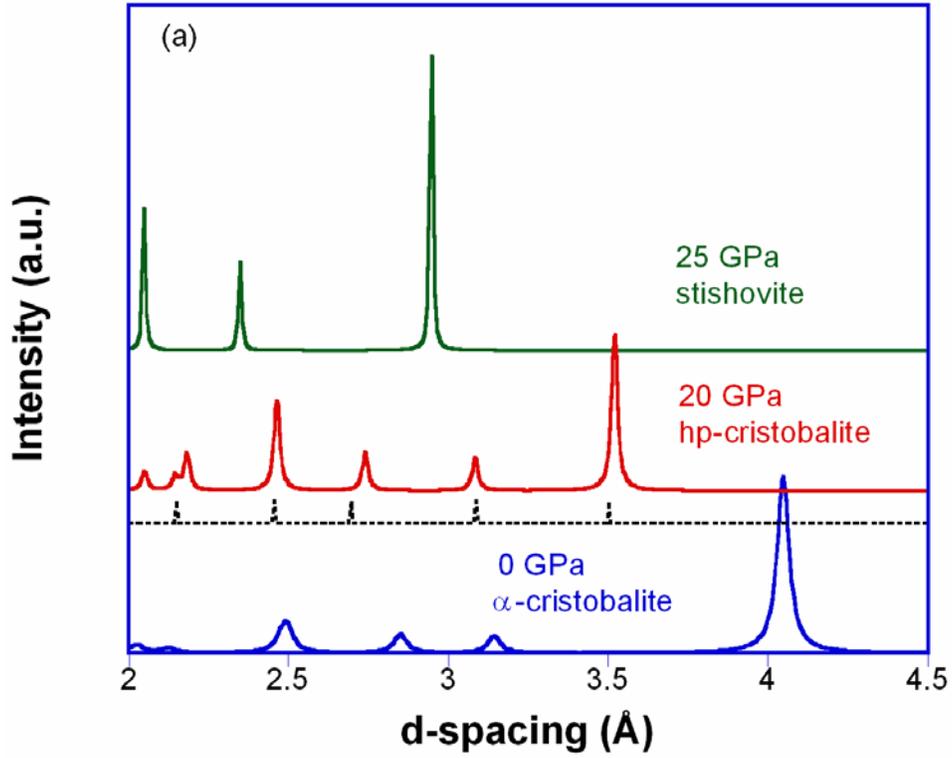

Fig. 2(a)-Huang-2005

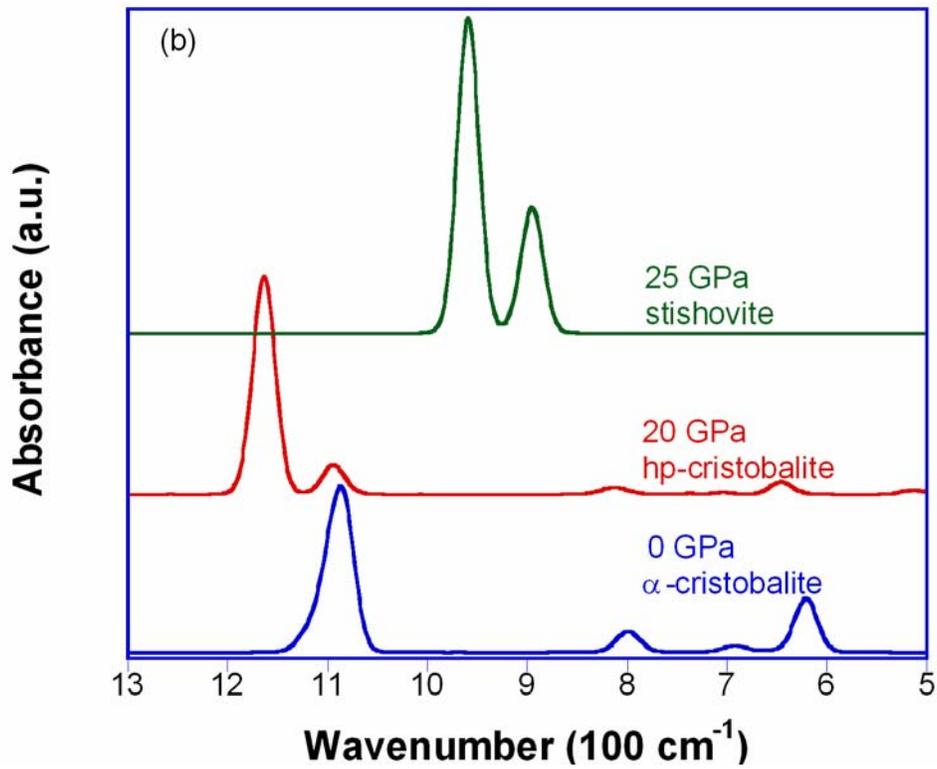

Fig. 2(b)-Huang-2005

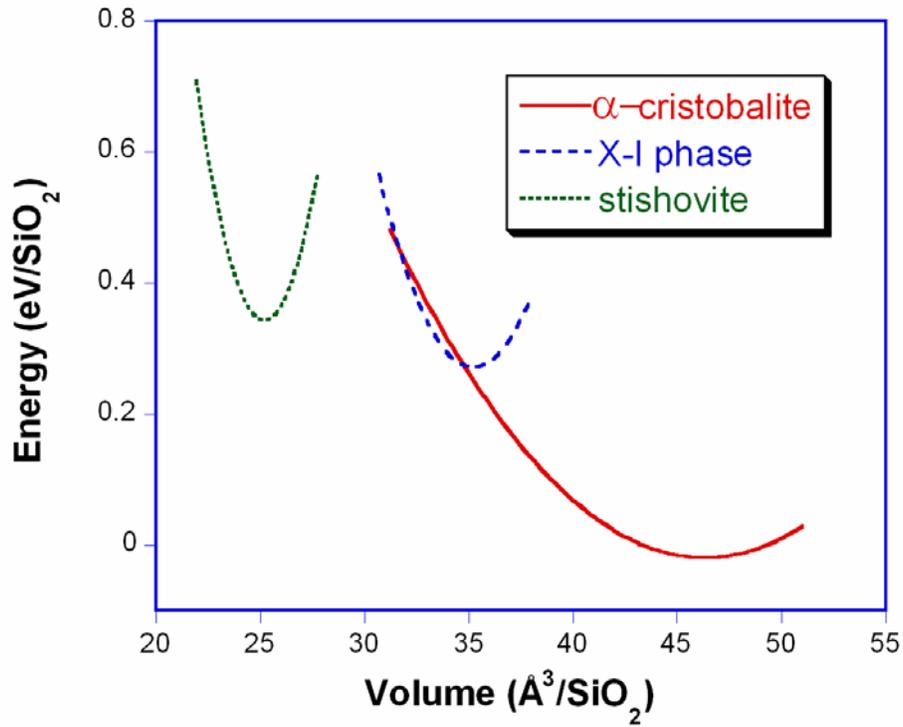

Fig. 3-Huang-2005

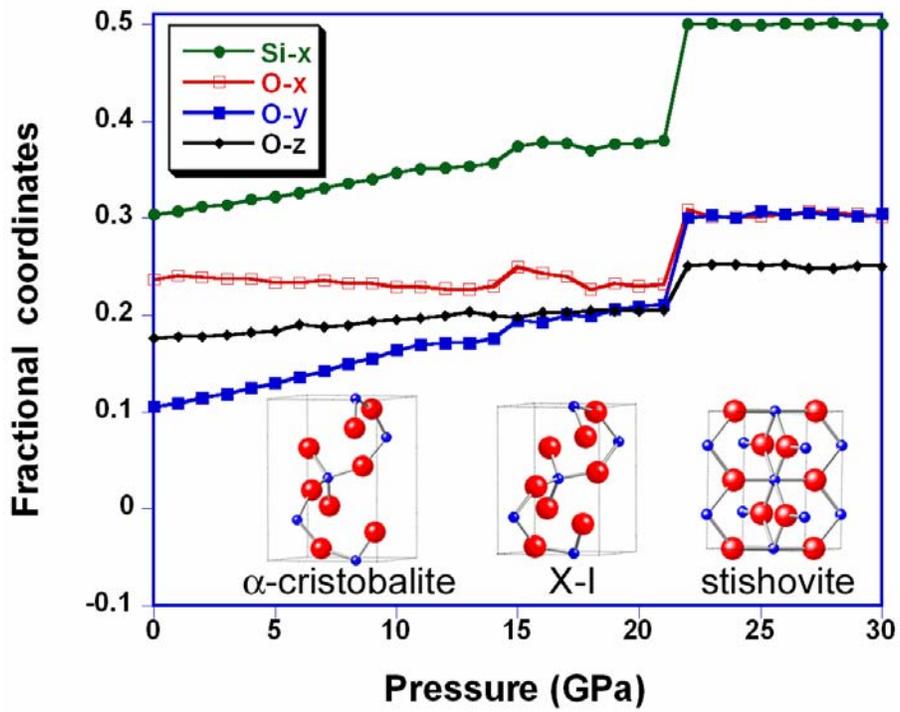

Fig. 4-Huang-2005

14